
\magnification=\magstep1
\baselineskip=20 truept
\def\ni{\noindent}
\def\hb{\hfil\break}
\def\bn{\bigskip\noindent}
\def\sn{\smallskip\noindent}
\def\pa{\partial}
\def\half{{\textstyle{1\over2}}}

\def\CC{{\bf C}}  
\def\fs{\vert f\vert^2}
\def\vp{\varphi}
\def\ve{\varepsilon}
\def\zb{{\bar z}}
\def\JI{J^{-1}}
\def\ms{|\mu|^2}
\def\xs{|\xi|^2}
\def\hb{\hfil\break}
\def\sech{\mathop{\rm sech}\nolimits}
\nopagenumbers
\font\scap=cmcsc10
\font\title=cmbx10 scaled\magstep1

\rightline{DTP95/57}
\vskip 1truein
\centerline{\bf Conserved Quantities for Integrable Chiral}
\centerline{\bf Equations in 2+1 Dimensions.}

\vskip 1truein
\centerline{\scap T. Ioannidou and R. S. Ward}

\bn\centerline{\sl Dept of Mathematical Sciences, University of Durham,}
\centerline{\sl Durham DH1 3LE, UK.}

\vskip 2truein
\ni{\bf Abstract.} The integrable (2+1)-dimensional chiral equations are
related to the self-dual Yang-Mills equation. Previously-known
nonlocal conservation laws do not yield finite conserved charges,
because the relevant spatial integrals diverge. We exhibit infinite sequences
of conserved quantities that do exist, and have a simple explicit form.

\bn{\it To appear in Physics Letters A.}

\vfill\eject

\pageno=2
\footline={\hss\tenrm\folio\hss}

\ni{\bf1. Introduction.}

\ni This paper deals with integrable chiral equations in 2+1 dimensions.
Locally, these are equivalent to the self-dual Yang-Mills (SDYM) equation,
reduced from 2+2 to 2+1 dimensions. However, the issue of boundary
conditions is crucial to our discussion, and we shall be using boundary
conditions which are natural for the chiral-equation form, rather than
the gauge-theory form.

The chiral equations in question have many of the
properties of an integrable system. For example, they arise as the
consistency condition for a pair of linear equations, and this description can
be used to generate multi-soliton solutions [1--5]; an inverse-scattering
transform can be set up [6, 7]; the equations satisfy the
Painlev\'e property [8]. A stricter characterization of integrability involves
the existence of sufficiently many conserved quantities in involution, and
hence a description in terms of action-angle variables. But for the integrable
chiral equations (and for SDYM) such an infinite set of conserved quantities
is not known.

Infinite sequences of conservation laws involving nonlocal densities
have been known for some time [9, 10]. These do not necessarily yield
conserved quantities, since the relevant spatial integrals may diverge.
For our chiral
equations, the solitons are localized in space, but the localization is
only polynomial, and the conserved densities referred to above do not fall
off fast enough at spatial infinity to give well-defined conserved quantities.

In this note, we demonstrate the existence of infinite sequences of
well-defined conserved quantities for the integrable chiral equations.
We also discuss local conserved quantities, including Noether charges
arising from symmetries of the Lagrangian. For simplicity, we take the
gauge group to be SU(2); but most of the discussion (and in particular the
infinite sequences of conserved quantities) extends automatically to
arbitrary gauge group.


\bn{\bf2. Integrable Chiral Equations.}

\ni The self-dual Yang-Mills equation in (2+2)-dimensional flat space can
be written in chiral-model form ({\it cf.\/} refs 9, 10). Dimensional
reduction then yields an
integrable chiral equation in 2+1 dimensions. But there is more than one way
of reducing, since the original chiral
equation in 2+2 dimensions is not SO(2,2)-invariant. The reduced equation
involves a choice of unit vector $V_\alpha$, and has the form
$$
 \pa^\mu(\JI J_\mu)-\half V_\alpha \ve^{\alpha\mu\nu}
       [\JI J_\mu, \JI J_\nu ] = 0. \eqno(1)
$$
Here $x^\mu = (t,x,y)$ are the space-time coordinates, $J$ is a $2\times2$
matrix of functions of $x^\mu$ with $\det J=1$, $\pa_\mu=\pa/\pa x^\mu$ and
$J_\mu=\pa_\mu J$ denote partial derivatives, $\ve^{\alpha\mu\nu}$ is
the totally skew tensor with $\ve^{012}=1$, and indices are raised and lowered
using the metric diag(-1,1,1). The vector $V_\alpha$ is constant, and
satisfies $V_\alpha V^\alpha = 1$; a Painlev\'e analysis suggests that (1)
is integrable if and only if $V_\alpha$ is such a unit vector [8].

If there were no further condition on $J$, then solutions of
(1) would correspond to the
gauge group SL(2,C). To reduce the gauge group to SU(2), we need to
impose a reality condition on $J$, the precise nature of which depends on the
choice of $V_\alpha$. The two cases we shall deal with are as follows.
\item{$\bullet$} Take $V_\alpha$ to be spacelike, specifically
 $V_\alpha=(0,1,0)$; and require $J$ to be unitary,
 {\sl ie.}\ $J\in{\rm SU(2)}$.
 With $u=\half(t+y)$ and $v=\half(t-y)$, eqn (1) can be rewritten as
 $$
  S:\qquad  \pa_v(\JI J_u)-\pa_x(\JI J_x)=0. \eqno(2)
 $$
\item{$\bullet$} Take $V_\alpha$ to be $i$ times a timelike unit vector,
 specifically $V_\alpha=(-i, 0, 0)$; and require $J$ to be hermitian
 (with positive eigenvalues). Putting $z=\half(x+i y)$, we can write the
 resulting equation as
 $$
 T:\qquad  \pa_t(\JI J_t)-\pa_\zb(\JI J_z)=0. \eqno(3)
 $$

\ni These two equations appeared in [2] and [1] respectively. Note that each
is equivalent to its hermitian conjugate; in other words, the reality
condition on $J$ is consistent with the equation. We impose the boundary
condition
$$
  J=J_0 + J_1(\theta)r^{-1} + O(r^{-2}) \eqno(4)
$$
as $r\to\infty$, where $x+i y =r\exp(i\theta)$. Here $J_0$ is a
constant matrix, and $J_1$ depends only on $\theta$ (not on $t$). This
condition  allows the existence of finite-energy soliton solutions.

The equation (1) has a global symmetry, in that $J$ can be multiplied on both
sides by constant matrices. If we require the reality conditions to be
preserved, then this global symmetry is SO(4) in the case of the $S$-equation
(2), and SO(1,3) in the case of the $T$-equation (3).

Finally in this section, let us remark on the static solutions of (2) and (3).
Static solutions of (2) are harmonic maps from ${\bf R^2}$ into SU(2).
Given the boundary condition (4), these are all known [11]: up to the
global symmetry mentioned above, they are
$$
  J={i\over1+\fs}\pmatrix{1-\fs &2f \cr
                            2\bar f &\fs-1 \cr}, \eqno(5)
$$
where $f$ is a rational function of either $z$ or $\zb$. By contrast, there are
no static soliton solutions of the $T$-equation (3). To see this, note that
(3) with $J_t=0$ implies that $\JI J_z$ is a holomorphic entire function
on $\CC$, tending to zero at infinity; hence $\JI J_z=0$ by Liouville's
theorem. But this means that $J$ is antiholomorphic entire, so $J=J_0$
constant, again by Liouville.


\bn{\bf3. Lagrangian and Local Conserved Quantities.}

\ni The classical conserved quantities arise, via Noether's theorem, from
symmetries of a Lagrangian. There is a Lagrangian for the integrable chiral
equation, although it involves breaking the global symmetry [10]. Let us
concentrate on the $T$-equation (3) for the time being, to see the
explicit expressions.

Following  well-established technique, we parametrize $J$ by
$$
  J=\vp^{-1}\pmatrix{1     &\bar\rho \cr
                     \rho  &\vp^2+|\rho|^2 \cr}. \eqno(6)
$$
Note that $\vp$ is well-defined and real, owing to the positivity of $J$.
Then (3) is equivalent to the Euler-Lagrange equations for the
Lagrangian
$$
  L=\vp^{-2}(\vp_t^2-|\vp_z|^2 + |\rho_t|^2 - |\rho_z|^2). \eqno(7)
$$
An obvious symmetry is the four-parameter family
$$
  \vp\mapsto |b|\vp, \qquad \rho\mapsto b\rho + c, \eqno(8)
$$
where $b$ and $c$ are complex constants. This is part of the global
symmetry noted previously. The corresponding conserved Noether densities
are components of $\JI J_t$. In fact, it is already clear that $\JI J_t$
is a matrix of conserved densities, since (3) has the form
of a conservation law. However, these densities go like $O(r^{-2})$ as
$r\to\infty$, and so the corresponding charges are not, in general,
well-defined.

The next obvious symmetries of (7) are the space-time translations, and
these lead to conserved energy-momentum. The energy density is
$$
  P^0 = \vp^{-2}(\vp_t^2 + |\rho_t|^2 + |\vp_z|^2 + |\rho_z|^2), \eqno(9)
$$
which is $O(r^{-4})$ as $r\to\infty$. So the energy $E$ (the spatial integral
of $P^0$) is a well-defined positive-definite functional of the field.
The momentum is also well-defined; the $x$-momentum density is
$$
 \eqalign{P^1 &= -\vp^{-2}(2\vp_t\vp_x + \bar\rho_t \rho_x
                          + \rho_t \bar\rho_x) \cr
              &= -{\rm tr}(\JI J_t \JI J_x), \cr}
$$
and the $y$-momentum density has the analogous form. Note that the momentum
is invariant under the full SO(1,3) global symmetry, whereas the energy
density is not (although it is invariant under the reduced symmetry (8)).

By way of example, let us examine the one-soliton solution [1]. This is
given by
$$
\eqalign{\vp &= {|\mu|(1+\fs)\over\ms+\fs}, \cr
         \rho&= {(\ms-1)\bar f\over\ms+\fs}, \cr}
$$
where $f$ is a rational meromorphic function of
$$
  \xi = \mu z + \mu^{-1}\zb - t,
$$
and $\mu$ is a complex constant with $|\mu|>1$. For the sake of simplicity,
take $f(\xi)=\xi$. Then
$$
 P^0 = {(\ms-1)^2(\ms+1) \over \ms(\ms+\xs)(1+\xs)}. \eqno(10)
$$
We see from (10) that the solution represents a single lump located at $\xi=0$.
This locus is a point in space which
moves in a straight line with constant speed $v=2|\mu|/(1+\ms)$.
The direction of motion is determined by the phase of $\mu$.
The energy of the soliton, expressed as a function of $v$, is
$E=8\pi\sech^{-1}v$. This has the \lq\lq anti-relativistic\rq\rq\
feature of decreasing as $v$ increases: $E\to0$ as $v\to1$ (the
speed of light), and $E\to\infty$ as $v\to0$, which is consistent with the
absence of  static solutions.

There is also a conserved angular momentum, corresponding to the rotation
symmetry $z\mapsto z\exp(i\chi)$. This, together with the energy and
momentum, are the only local conserved quantities of which we are aware; and
their existence is not really connected with the integrability of the equation.

In [12] there is a description of infinite sets of local conserved densities;
these are constructed from
differential operators acting $\JI J_\mu$, and their integrals diverge in
the same way as that of $\JI J_t$. Even if a
boundary condition were chosen which ensured convergence, integration by parts
gives relations between these conserved charges which seem to
indicate that very few, if any, of them are independent and new. In other
words, one still only has the obvious local conservation laws, irrespective of
convergence.

We conclude this section by mentioning the corresponding results for the $S$
equation~(2). It is evident from the equation that $\JI J_u$ is a
conserved density; but, as before, the corresponding charges diverge.
The energy and $y$-momentum are well-defined, their densities being given by
$$
\eqalign{P^0 &= -\half{\rm tr}\bigl[(\JI J_t)^2+(\JI J_x)^2
                               +(\JI J_y)^2\bigr], \cr
         P^2 &= {\rm tr}(\JI J_t \JI J_y). \cr }
$$
A conserved $x$-momentum density can be obtained from a Lagrangian
analogous to (7). But the functions appearing in it have singularities in
general (in particular, this is the case for the soliton solutions); and
as a consequence, the $x$-momentum is divergent. The problem occurs because
of singularities in a parametrization such as (6) in this case.
It is not obvious whether
one can find another parametrization which avoids these singularities.


\bn{\bf4. Nonlocal Conserved Quantities.}

\ni An infinite sequence of nonlocal conserved currents
was first exhibited by Prasad {\sl et al}\ [9], and independently by
Pohlmeyer [10]. This was motivated by an analogous sequence for the
two-dimensional chiral model. A different sequence of nonlocal currents
was later given by Leznov ({\sl cf.}\ [13]), and independently by
Papachristou [14]. A third  sequence was mentioned by Sutcliffe [4].

Leznov's argument made use of an alternative form of the SDYM and the chiral
equations. For the $S$-version (2), this is
$$
X_{xx}-X_{uv}+[X_v , X_x] = 0, \eqno(11)
$$
where $X$ is a Lie-algebra-valued function defined by
$X_x=\JI J_u$, $X_v=\JI J_x$.
Eqn (11) also arises from a Lagrangian [13], but the corresponding energy
functional is not positive-definite. Papachristou [15] called (11) the
\lq\lq potential SDYM equation\rq\rq. He pointed out that
new conservation laws could be derived from
symmetries of (11), and that such symmetries are in effect solutions of a
novel linear system for the SDYM equations [16]. This latter possibility
has yet to be fully explored.

For the time being, let us concentrate on the case of the $S$-equation (2).
The ``old'' nonlocal conserved densities involve the integral operator
$\pa_x^{-1}$, which we take to be $\pa_x^{-1} F(x)=\int_{-\infty}^xF(x')\,dx'$.
Write $A=\JI J_u$, $B=\JI J_x$, and $C=\pa_x^{-1}A$. Then the first
nonlocal conserved densities of [9, 10], and of [13, 14], are respectively
$$
\eqalign{ a &= B + 2 C_y + AC, \cr
          b &= B + 2 C_y + \half[A,C]. \cr
}$$
The boundary condition (4) implies that $A$ and $B$ are $O(r^{-2})$ as
$r\to\infty$ (so $C$ is bounded but nonzero at infinity). Consequently, the
spatial integrals $\int\!\!\!\int a \,dx\,dy$ and $\int\!\!\!\int b \,dx\,dy$
are divergent in general. The same is true for the other nonlocal conserved
densities in the sequences of [9,~10,~13,~14].

Now observe that $2(a-b)=AC+CA=\pa_x C^2$. Although the integral of this is
formally divergent, there is a way of defining it which makes sense. This gives
the first member $Q_1$ of our new sequence $\{Q_n\}$, which we now define.
The  $Q_n$ are given by
$$
  Q_n = \int_{-\infty}^{\infty} M^{n+1} \, dy, \quad{\rm for}\ n=1,2,\ldots
        \eqno(12)
$$
where
$$
  M(t,y) = \int_{-\infty}^{\infty} \JI J_u \, dx. \eqno(13)
$$

First, let us show that these $Q_n$ are well-defined, for any fixed value of
the time~$t$. The boundary condition implies that there exists a positive
constant~$K$ such that each component~$A_{\alpha\beta}$ of the
matrix~$A=\JI J_u$ satisfies $|A_{\alpha\beta}|\leq K/(r^2+1)$, and so
$|M_{\alpha\beta}|\leq\pi K/\sqrt{y^2+1}$. It follows that
$M^{n+1}=O(y^{-n-1})$ as $|y|\to\infty$, and so the integral~(12) converges.

Secondly, we prove that the~$Q_n$ are conserved. Note that
$$
\pa_v M = \int_{-\infty}^\infty \pa_v(\JI J_u)\,dx
        = \int_{-\infty}^\infty \pa_x(\JI J_x)\,dx=0.
$$
Hence $\pa_v M^{n+1}=0$, and so
$$
 {d Q_n\over dt} = \int_{-\infty}^\infty \pa_t M^{n+1}\,dy
        = \int_{-\infty}^\infty \pa_y M^{n+1}\,dy = 0,
$$
as claimed.

Thirdly, let us investigate how many independent conserved quantities there
are among the $Q_n$. This discussion applies specifically to the SU(2) case
(whereas the previous remarks apply equally well to any other gauge group).
Since $M$ takes values in the Lie algebra su(2), we have
$$\eqalign{
           M^{2p} &= (-1)^p \Vert M\Vert^{2p} I, \cr
         M^{2p+1} &= i(-1)^p \Vert M\Vert^{2p} M, \cr
}$$
where $I$ is the identity $2\times2$ matrix, and $\Vert M\Vert^{2}
=-\half{\rm tr} (M^2)$. So for each odd $n$, there is one real conserved
quantity ($Q_n$ equals, up to sign, the number $\int\Vert M\Vert^{n+1}\,dy$
times $I$); and for each even $n$, there are three (the components of
$\int\Vert M\Vert^n\,M \,dy$). Since $M$ is essentially an arbitrary
su(2)-valued function of $y$ (think of initial data for $J$ on $t=0$, say),
all these conserved quantities are independent.

There is another set of well-defined conserved charges, which are
complementary to, and independent of, the $Q_n$. They are related to the
(nonintegrable) conserved densities given in [4].
The key point is that (2) is equivalent to
$$
  \pa_u(J_v\JI) - \pa_x(J_x\JI) = 0.
$$
{}From this, it is easy to see that the following quantities are conserved:
$$
  \widehat Q_n = \int_{-\infty}^{\infty} \widehat M^{n+1} \, dy, \quad
               {\rm for}\ n=1,2,\ldots \eqno(14)
$$
where
$$
  \widehat M(t,y) = \int_{-\infty}^{\infty} J_v\JI \, dx. \eqno(15)
$$

For the $T$-equation (3), the situation is somewhat different.
Once more, the conserved densities of [9, 10] are not integrable;
in particular, this is the case for the soliton solution
described in section 3. So the corresponding conserved charges do not
exist. However, one can define charges analogous to the $Q_n$ above. These
are
$$
  R_n = \int_{{\bf R^2}} \pa_\zb (\Psi^n)\, dz\wedge d\zb, \eqno(16)
$$
where $\Psi$ is the solution of the $\bar\pa$-problem
$$
  \pa_\zb \Psi = \JI J_t, \qquad \Psi\to0\ {\rm as}\ |z|\to\infty. \eqno(17)
$$
The conservation of the $R_n$ follows from $\pa_t \Psi=\JI J_z$.

\bn{\bf5. Concluding Remarks.}

\ni We have exhibited several sequences of well-defined conserved
quantities for integrable (2+1)-dimensional chiral equations. Although
nonlocal, they have a particularly simple form, not involving repeated
integration. They are related to nonlocal conservation laws known
previously; however, the latter do not yield well-defined conserved
quantities, and so cannot contribute to ``a complete set of action
variables''.

The present sequences certainly do not make up a complete set:
for example, the
matrix $M$ vanishes for the one-soliton solution (5) and its
moving version, and therefore so does $Q_n$. So one needs to search for more
conserved quantities, and also to relate them
to the algebraic-geometry [3] and inverse-scattering [6,~7]
approaches to the chiral equations. In this regard, it may be useful to compare
with the sine-Gordon equation, which is a dimensional reduction of the
integrable chiral equations [17]; however, the comparison cannot be a direct
one, since the boundary conditions are different. In the case of sine-Gordon,
much more is known about conserved quantities, both local and nonlocal~[18].

\bn{\bf Acknowledgment}

\sn The first author thanks the Federation of West Germany Pontium for
financial support.

\vfil\eject

\ni{\bf References.}

\ni[1] S. V. Manakov and V. E. Zakharov, Lett. Math. Phys. 5 (1981) 247.\hb
\ni[2] R. S. Ward, J. Math. Phys. 29 (1988) 386.\hb
\ni[3] R. S. Ward, Commun. Math. Phys. 128 (1990) 319.\hb
\ni[4] P. M. Sutcliffe, Phys. Rev. D 47 (1993) 5470.\hb
\ni[5] P. M. Sutcliffe, J. Math. Phys. 33 (1992) 2269.\hb
\ni[6] J. Villarroel, Inverse Prob. 5 (1989) 1157.\hb
\ni[7] J. Villarroel, Stud. Appl. Math. 83 (1990) 211.\hb
\ni[8] R. S. Ward, Nonlinearity 1 (1988) 671.\hb
\ni[9] M. K. Prasad, A. Sinha and L.-L. Wang, Phys. Lett. B 87 (1979) 237.\hb
\ni[10] K. Pohlmeyer, Commun. Math. Phys. 72 (1980) 37.\hb
\ni[11] K. K. Uhlenbeck, J. Diff. Geom. 30 (1989) 1.\hb
\ni[12] C. J. Papachristou and B. K. Harrison, Phys. Lett. A 127 (1988) 167.\hb
\ni[13] A. N. Leznov and M. V. Saveliev, Acta Applic. Math. 16 (1989) 1.\hb
\ni[14] C. J. Papachristou, Phys. Lett. A 138 (1989) 493.\hb
\ni[15] C. J. Papachristou, Phys. Lett. A 145 (1990) 250.\hb
\ni[16] C. J. Papachristou, J. Phys. A 24 (1991) L1051.\hb
\ni[17] R. A. Leese, J. Math. Phys. 30 (1989) 2072.\hb
\ni[18] R. Sasaki and R. K. Bullough, Proc. R. Soc. Lond. A 376 (1981) 401.

\bye